\documentclass{article}   
\usepackage{natbib}
\usepackage[nottoc,numbib]{tocbibind}

\usepackage{physics,amsmath,amssymb,mathrsfs}
\usepackage{enumitem}

\title{A (strictly) contemporary perspective on trans-Planckian censorship\footnote{Accepted at \emph{Foundations of Physics}}}
\author{Mike D. Schneider\footnote{University of Illinois at Chicago}}
\date{}

\begin{document}
\maketitle

\begin{abstract}
I critically discuss a controversial `trans-Planckian censorship' conjecture, which has recently been introduced to researchers working at the intersection of fundamental physics and cosmology. My focus explicitly avoids any appeals to contingent research within string theory (the sociological origins of the conjecture) or regarding the more general (quantum) gravitational `swampland'. Rather, I concern myself with the conjecture's foundations in our current, well-trodden physics of quantized fields, spacetime, and (classical) gravity. In doing so, I locate what exactly within trans-Planckian censorship amounts to a departure from current physics --- identifying what is, ultimately, so conjectural about the conjecture.
\end{abstract}

\tableofcontents

\section{Introduction}

The `trans-Planckian censorship' conjecture (TCC) was first introduced to the physics preprint repository ArXiv in a pair of articles posted in September 2019, later published as \citep{bedroya2019trans} and \citep{bedroya2020trans}. More recently, the TCC has been advertised --- by the first author on both of those inaugural articles --- as ``a  natural  modification  of  the  renormalizability condition for gravitational theories where the conventional notion of renormalizability does not apply'' \citep[p. 6]{bedroya2020sitter}. This is quite an advertisement! Renormalization in quantum field theory is ``one of those great success stories in physics'' \citep[p. 1]{rivat2019renormalization}. Indeed, it is often taken as a condition of adequacy in the construction of fundamental physical theories within quantum field theory that such constructions \emph{be renormalizable}. So too may one read the present claim as an insistence that the TCC will, in the future, be rendered true as a condition of adequacy in theory construction within the context of quantum gravity (i.e. `beyond field theory'). 

Meanwhile, another author from one of those inaugural articles has described the TCC as ``a momentum space generalization of Penrose's [cosmic censorship] hypothesis'' familiar from general relativity \citep[p. 6]{brandenberger2021limitations}. And lest one think that this second advertisement is subdued in comparison with the first: I can report that in a discussion section held on the second day of the inaugural meeting of the International Society for Quantum Gravity (organized virtually in October 2021), the same author further offered the TCC, on this second advertisement, as an organizing principle behind the entire (quantum) gravitational `swampland' program.

The latter is a momentous claim, and certainly on par with the claim of the first advertisement above. Originally developed in the context of studying obstructions to stringy ultraviolet (UV) completions of effective field theories (EFTs), the swampland program now describes a more general attempt to study which quantum field theoretic constructions interpreted as low-energy EFTs cannot describe material quantum systems that are (quantum) gravitationally coupled \citep{palti2019swampland}. Research in the swampland program proceeds in terms of `swampland conjectures'. Many are narrowly focused technical conjectures about families of EFTs that safely avoid the swampland, or otherwise fall within it: failing to admit UV completions, given (quantum) gravity. These technical conjectures are motivated directly by arguments and toy models in particular UV quantum gravity approaches --- traditionally string theory. But some swampland conjectures are quite broad. For instance, string universality is one such broad conjecture, which states that the string `lamppost principle' is true: EFTs that avoid the swampland (that is, those admitting UV quantum gravity completions) admit stringy UV completions \citep{van2021lectures}. Swampland arguments in support of string universality are therefore arguments in support of a sustained interest in a string theory approach to quantum gravity (cf. \citep{kim2020four}). It is in this sense that swampland research, though originating in string theory, would seem to represent a strictly more general --- perhaps even `approach-neutral' --- means of conducting quantum gravity research. So an assertion that the TCC is an organizing principle behind the swampland program would seem to amount to a far reaching and profound claim about the conjecture's place within the overall conceptual structure of the problem of quantum gravity.  

What is this incredible new conjecture? Roughly speaking, the TCC prohibits `trans-Planckian' quantum modes from becoming `trans-Hubble' in any EFT treatment of a (gravitationally coupled) material quantum system described quantum field theoretically (i.e. described as a quantized relativistic field). That is, considered in terms of an EFT, any initially UV, super-Planckian modes indexed by frequency in `momentum space' that dynamically stretch, in the course of a background rapid expansion of spacetime due to gravity, to sub-Planckian frequency scales are prohibited from further stretching to scales on the order of the (inverse) Hubble radius --- a rough lower bound on the characteristic scale of classical physical interactions relevant in a linearized `momentum space' treatment of the growth of classical structure in large-scale cosmology. 

But as I discuss in \citep{selfreferenceTransPlanckianPhilosophy}, the normative force of any physical consequence of this prohibition is, by the conjecture, secured by unknown facts specifically about UV quantum gravity. The TCC is an assertion about what is possible of quantum field theoretic physics, as interpreted by means of EFTs, given some as-yet unknown UV theory of quantum gravity that takes over as an apt description of the same material quantum systems in our gravitationally coupled universe at energies above the Planck scale (the scale above which it is standard to expect that any field theoretic description of a gravitationally coupled system will simply fail to be apt). Since we do not yet have that UV theory of quantum gravity, the upshot is that whether one is to presently embrace the conjecture simply must amount to speculation about future physics. 

And as it happens, one physical consequence of this speculation is a derivation of an upper bound in fundamental physics on the duration of the background rapid expansion in any EFT description of a material quantum system in our gravitationally coupled universe. As originally noted by \citet{bedroya2020trans}, this would seem to apply significant pressure on the popular `inflationary cosmology' research program in early universe cosmology (thereby motivating work on alternatives to inflationary accounts of large-scale cosmic structure formation),\footnote{The upper bound on the duration of rapid expansion allowed (e.g. in an inflationary epoch in the early universe) depends on fixing a class of models, given further empirical constraints (e.g. the amplitude of the CMB power spectrum downstream of the inflationary epoch). On this point, as summarized in \citep[p. 7]{brandenberger2021limitations}, bounds on inflation that are due to trans-Planckian censorship:\begin{quote}... are only consistent if inflation takes place
at a very low energy scale of $V^{\frac{1}{4}}< 10^9$GeV, assuming an almost constant value of [Hubble radius] during inflation and instantaneous reheating. This is a very severe constraint on canonical single scalar field models of inflation. The bound can be relaxed by allowing a substantial time dependence
of [Hubble radius] during inflation or by modifying the post-inflation cosmology. On the other hand, the bound is strengthened if there is a period of radiation domination before the onset of inflation.\end{quote}} as well as on one popular understanding in standard model cosmology of dark energy in the asymptotic future (cf. \citep{heisenberg2018dark}). Unsurprisingly then, the TCC has, as speculation, been met with considerable resistance in the wider community --- this, despite the spectacular advertisements that have been put forth in its favor. On the other hand, the speculation has also been used for certain \emph{further} ends in cosmology, including late-stage cosmology beyond the standard model. For instance, in the new program in cyclic cosmology proposed in \citep{ijjas2019new}, one might argue that trans-Planckian censorship is implicitly invoked in order to help derive a finite upper bound on the lifetime of any given cycle.\footnote{\label{fnCyclic}Or, more carefully put: a principle along the lines of trans-Planckian censorship is evidently what would ensure, in a model-independent way, that there is a finite upper bound in each cycle --- so that, especially, it might thus be reasonable to go `beyond the standard model' in late-stage cosmology, favoring end conditions consistent with a cyclic model instead. (In \S3.1 of the paper cited in text, the model included for illustrative purposes builds in the finiteness by hand, though the authors do explicitly note that facts about the dark energy density are what, in the model, matters to that finiteness.) I thank Anna Ijjas for correspondence on this topic.}

In light of the conjecture's ambivalent reception, it is of some importance that one look beneath the spectacular advertisements for the TCC. These advertisements are, after all, offered in the contingent thicket of ongoing quantum gravity research. By contrast, what one would really like is to better understand arguments that might motivate the speculation, given just our current, well-trodden physics of quantized fields, spacetime, and (classical) gravity. And this requires that great care be given, in the first place, to how one presents the TCC just in terms of that current, established physics. 

Unfortunately, to my knowledge, such a presentation has not yet been provided. In fact, the existing sentiment in the literature would seem to favor a view that is very nearly the opposite: that, in light of the pressures it places on inflationary cosmology, what one is after in order to motivate the TCC is an argument entirely `within' quantum gravity research. For instance, in one of the clearest articulations of this sentiment that I have found, \citet[p. 5]{brahma2020trans} writes that:\begin{quote}
If correct, the TCC would imply a tectonic shift in our understanding of early-universe cosmology since it seems to highly disfavour models of inflation. Given its radical consequences, it is natural to ask if the TCC can be obtained from some other well-tested aspect of quantum gravity.\end{quote} Here, `well-tested' is to be understood entirely as a matter of theoretical work in the course of ongoing quantum gravity research --- not to do with any particular physical theory's purchase on empirical phenomena. But I think this confuses ease of argument exposition for argumentative strength. A stronger argument is one that is ultimately \emph{unconditional} on (plausibly) contingent claims arising within quantum gravity research. Better, then, is to proceed directly from a firm empirical foundations provided by our current best theories.

Consequently, my primary goal in this article is to supply such a presentation: developing foundations for the TCC, as far as is possible, just by means of current physics. The upshot of my doing so will be, along the way, to clarify several interrelated points regarding the conceptual and logical structure of trans-Planckian censorship, as it is situated at the nexus of fundamental physics and cosmology. These points include the relationship between the TCC and the notion of limits on the use of the EFT toolkit in fundamental physics; the role of classical, large-scale cosmology in motivating the TCC; and how (and how not) speculating the TCC bears relevance to ongoing \emph{semiclassical} cosmological modeling practices. 

Through these means, by the end of the article I will have located what exactly within an affirmation of trans-Planckian censorship would amount to a departure from our current physics --- locating what, exactly, is so conjectural about the conjecture. I will conclude that the TCC is speculative in virtue of its invoking a particular new fundamental physical claim, entirely beyond the scope of current physics. Namely (in terms that will be made abundantly clear below): effective field theoretic descriptions of gravitationally coupled material quantum systems in the cosmological sector of UV quantum gravity must feature a low-energy (`infrared', or `IR') cutoff --- a cutoff that is, in particular, specifically tailored to classical considerations in cosmological modeling. Reflecting on how, exactly, this particular `UV/IR mixing' claim extends beyond our current physics applies at least some pressure on the persuasiveness of the above advertisements for the TCC, which might otherwise seem to cast our speculating trans-Planckian censorship as an eminently compelling prospect within ongoing quantum gravity research. 

\section{The anatomy of the TCC}

The goal of this section is to flesh out an understanding of trans-Planckian censorship, as one may presently articulate such conjectural physics. I do so in three installments --- \S2.1, 2.2, and 2.3 respectively. This paves the way for discussion in \S3 as to whether the TCC might then be understood, from the perspective of just our current physics, to perform some crucial role in the context of UV quantum gravity. 

Of course, on this upshot: it is important to not lose sight of the fact that the TCC is ultimately a claim that is made (by conjecture) in accordance with an as-yet unknown UV theory of quantum gravity. It is not \emph{about} our current, well-established fundamental physical theories --- i.e. general relativity (GR) and quantum field theory (QFT). As is well known, the problem in fundamental physics that research toward a UV theory of quantum gravity is meant to address begins with recognition of a tension between GR and QFT (only then bolstered by further physical arguments that draw on additional conceptual resources, e.g. those of statistical mechanics). Namely, descriptions of a real-world --- hence, gravitationally coupled --- material quantum system that are owed to GR are, on a flat-footed reading, in contradiction with descriptions of the same system that are owed to QFT. As typically presented, these contradictions reflect the fact that the former descriptions are entirely classical, despite the material system being quantum; hence, one arrives at the usual call for some future quantum theory of gravity to ultimately replace GR.\footnote{In this exposition, I am being purposefully quiet about whence the demand for a quantum gravity theory that is, specifically, `UV'. It suffices to say that, in the wake of GR, there are conceptual challenges spelling out adequacy criteria for \emph{any} fundamental quantum gravity theory, if that theory is not apt for use above the Planck scale. But this subtlety is a tangent in the present context.} Thus, one should not be surprised to note that our current theories of GR and QFT will, in themselves, be inadequate to the task of substantiating the current physical foundations of the TCC.

Instead, in contemplating consequences of a UV theory of quantum gravity that is yet to come, we must all content ourselves with `semiclassical gravity'. Semiclassical gravity is a theory (more or less) that amalgamates our current physical commitments stemming, respectively, from GR and QFT, as each would individually be applied to some given system. As I discuss in \citep{schneider2020s}, the theoretical architecture of semiclassical gravity (based, as it is, on our current, well-established physical theories) is intended to provide a means of generating approximations of the low-energy consequences of the future UV theory of quantum gravity in application to the given system --- that is, given our embrace of GR and QFT as incredibly successful theories that preside over particular empirical domains. Meanwhile, taking semiclassical gravity --- in light of those successes --- to provide such approximations amounts to an anticipation, today, about that future physics (in the sense of \citep{schneider2020betting}). So, any claims made about the real-world system on the basis of semiclassical gravity are ``not consequences of current physical theory. Rather, they are claims about what one might infer from contemporary physics about approximations of future physics'' \citep[p. 15]{schneider2020s}. Just so, trans-Planckian censorship is a conjecture about future physics, and the task at present is to articulate the conjectured claim \emph{as well as is presently possible}: by way of an (anticipated) approximation provided by the theoretical architecture of semiclassical gravity.

Recently, \citet{Wallace} has argued forcefully for the point of view that the theory `low-energy quantum gravity' belongs alongside those EFTs familiar from standard model particle physics in the corpus of empirically successful theories that together characterize our current best fundamental physics. Here, low-energy quantum gravity refers to an interacting EFT of the `graviton', constructed bottom-up from (linearized) perurbative GR about any given background --- where the background is then interpreted as a mean-field expression appropriate at low-energies as an approximation of some given state in an underlying unknown UV quantum gravity theory. As \citet{Wallace} notes, low-energy quantum gravity can thereby lay claim to the `semiclassical Einstein' equation that is otherwise characteristic of the dynamics associated with semiclassical gravity: now interpreted as a part of the mean-field description against which effective fluctuations in low-energy quantum gravity are to be perturbatively modeled. Hence, within the context of low-energy quantum gravity, the theoretical architecture of semiclassical gravity arguably captures \emph{more than approximations of future physics}: yielding instead genuine consequences of current physical theory (entirely analogous to those cases of QFT that are relevant in standard model particle physics).

At first glance, this argument would seem to contradict the view of semiclassical gravity as, ultimately, a (mere) amalgamation of our empirically successful theories GR and QFT, intended to anticipate approximations of future physics. But this is not so. Indeed, I am happy to accept \citeauthor{Wallace}'s argument that the semiclassical Einstein equation arises as a part of our current physics in exactly the way he outlines, concerning the theory of low-energy quantum gravity. And so, I do not wish to contest his conclusion that consequences of the application of the semiclassical Einstein equation in low energy quantum gravity may express genuine statements about physics, as according to current theory. Still, I maintain that there is a \emph{wider} conception of semiclassical gravity than just the mean-field theory of low-energy quantum gravity. And in that wider conception, applications of the semiclassical Einstein equation are rather to be understood as above: namely, as providing anticipated approximations of future physics. Moreover, this conception remains relevant to contemporary fundamental physics research, so long as the classical theory GR --- or even just what \citet{Wallace} clarifies as `phenomenological GR' --- is understood to keep its place at the fundamental physicist's table.

And indeed, there is good reason to want the classical theory to keep its place at the table --- perhaps especially if low-energy quantum gravity has earned a place, as well. As will become evident, the TCC is a UV quantum gravity conjecture that specifically places domain boundaries on the scope of low-energy quantum gravity --- that is, boundaries in its application to real-world systems, which are otherwise ambiguous in light of our lack of understanding about the allowed states in the underlying UV quantum gravity theory. But for one to presently articulate those boundaries, as a matter of approximation of some better articulation yet to come, one clearly requires that the corpus of our current best fundamental physics includes \emph{more than just EFTs}. In this respect, the TCC draws on more of current physics than just what is invoked in  low-energy quantum gravity: it will not suffice to consider only the interactions between an effective, quantized field and gravitons (i.e. about a shared, fixed background that is assumed to satisfy mean-field theory equations).\footnote{This is, in fact, a subtle point about the physics implicated in trans-Planckian censorship. At face value, the statement of the TCC only involves reference to that part of the physics circumscribed by low-energy quantum gravity (which is, after all, an effective theory defined perturbatively, relative to any suitable choice of background, including that which will feature shortly in \S2.1). But as evidenced by the later discussion in \S2.3, there is more physical reasoning involved than just the effective field theoretic dynamics of material field modes interacting with low-energy quantum gravity, situated on a background.}

This observation about the current physics implicated in trans-Planckian censorship is, I take it, much in the spirit of \citep{koberinskiSmeenkRethinkingCCP}. In that article, the authors argue that the cosmological constant problem might best be understood as a reductio on the application of EFTs to fundamental physical modeling in large-scale cosmology. As is apparent in their article, semiclassical gravity --- in particular (and contra \citeauthor{Wallace}), a conception of semiclassical gravity \emph{large enough to incorporate developments in quantum field theory on curved spacetime (QFTCS)} --- provides the means to make such an argument. 

And so too is the framework of QFTCS, situated within the context of semiclassical gravity, crucial for trans-Planckian censorship. Recall from the Introduction that the TCC was summarized as a prohibition, in the context of an EFT, on trans-Planckian quantum modes dynamically stretching into trans-Hubble modes. It is precisely the architecture of QFTCS that can underpin such claims to do with `dynamical stretching' of modes in a real-word (i.e. gravitationally coupled) material quantum system: in particular, where the curved spacetime --- encoding the properties of the relevant classical gravitational state according to GR --- happens to be a spatially `expanding' one. Elaborating on this point will provide a natural place to begin a strictly contemporary anatomy of the conjecture. (It is also, I will note in advance, the most substantial contribution to that anatomy, of the three contributions I will name.)

\subsection{Dynamically stretching modes}\label{secDynamicallystretching}

In QFT, it is taken for granted that material quantum systems are modeled as quantized special relativistic fields --- fields defined over the special relativistic `Minkowski' spacetime --- which may be prepared in arbitrary quantum states. Notably, gravity is here conceptualized entirely in terms of a `weak-field regime' of GR: as local metric perturbations on the Minkowski background, so that in most applications that are localized to particular events in spacetime, gravity may be safely ignored. Namely, about the given events, gravitational states are understood in terms of local curvature, which in units relevant to any given modeling context most often approximately vanish.

But according to GR, spacetime is, in a gravitationally coupled material system, more aptly represented by members of a whole class of Lorentzian manifolds --- hence, the ``general'' in ``general relativity'' \citep{malament2012topics}. Thus, the \emph{global} geometric properties of the particular members in that class are what are understood to encode viable descriptions of gravitational states, as may be associated with some or other material system. Alternatively, treating so-called `global spacetime structure' as roughly (that is to say, entirely informally) analogous to superselection sectors in a quantum theory,\footnote{See \citep{earman2008superselection} for a classic foundational treatment of the latter. Note that, in this rough analogy, it is very likely not the case that the underlying topological and smoothness structure of a spacetime corresponds to a global charge. Causally perverse spacetimes defined on trivial underlying smooth manifolds, e.g. G{\"o}del spacetime, complicate the story \citep{stein1970paradoxical}. If, though, we are restricted to globally hyperbolic spacetimes (as discussed presently), it seems plausible that (smoothly) foliated manifolds are what may stand in that rough analogy.} the \emph{local} geometric properties of those same members, specified in terms of metrics defined on underlying smooth manifolds, encode viable configurations of the classical gravitational \emph{field}. (And from this perspective, the reliance of QFT on the weak-field regime of GR depends on the suitability of something analogous to a superselection rule over classical gravitational states.) 

QFTCS aims, in the first place, to provide descriptions of the behaviors of material quantum systems as quantized relativistic fields, given this latter conceptualization of classical gravity in the background --- that is, a conceptualization that does not \emph{presuppose in every ensuing successful application} that such a success reflects our being safely within a weak-field regime of GR. 

Here, I primarily follow \citet{wald1994quantum}, though my emphasis on foundations will lead me to a slightly different presentation than that found in his pedagogical text. (See also \citet[\S{3-4}]{koberinskiSmeenkRethinkingCCP}, including favorable references therein that are more recent than the 1990s.) Of course, how one delimits spatiotemporal possibility in the classical theory GR may shrink or enlarge the scope of QFTCS in fundamental physics. Usually, one's choices in how to define the class of Lorentzian manifolds that are relevant reflects anticipations about the UV theory of quantum gravity to come. For instance, restricting attention to those members in the class that satisfy cosmic censorship may be taken as a commitment to UV quantum gravity resolving (whether by fiat or by some further explanation) the problem of naked singularities in the state space of the classical theory recovered in a suitable limit. The strongest version of cosmic censorship mandates that spacetime be globally hyperbolic. In this case, spacetime is diffeomorphic to a product manifold $\mathbb{R}\times S$, with $S$ a (smooth) spacelike Cauchy surface \citep{bernal2003smooth}; indeed, at least for non-interacting quantized fields, we know how to proceed for arbitrary globally hyperbolic backgrounds, taking advantage of this product structure for the diffeomorphism class of any such gravitational state.

From this perspective, as a special project of interest, it is an aim of QFTCS to provide descriptions of classical gravitationally coupled quantized fields: (globally hyperbolic) spacetime backgrounds classical gravitationally sourced by material quantum systems modeled as quantized fields \emph{prepared in arbitrary states}. Unfortunately, this special project is exceedingly difficult, if even remotely viable as posed. Still, the statement of the project directs our attention to a more basic question of how to introduce a classical gravitational coupling into our physical descriptions of the material quantum system, modeled as a quantized field. This is where QFTCS becomes a framework for doing semiclassical gravity.\footnote{Another avenue of research in QFTCS sets aside semiclassical gravity, in order to focus on the phenomenology of quantized fields merely subject to specific model spacetime backgrounds selected for their applications \citep{birrell1984quantum}. But this will not suffice for our purposes: as will become clear below, it is crucial to the TCC that we consider the material quantum field on the expanding background as gravitationally coupled, such that expansion is itself an apt mean-field description of the gravitational state of the total system, including small perturbations.} Starting from the semiclassical Einstein equation, which is directly inspired by the phenomenological successes of GR, one concludes that a certain `stress-energy tensor' operator is required in the QFT on top. This is an operator whose action on a freely given state of the quantized field is, in the case of a two-point function in a coincidence limit about a point, determined point by point in the curved spacetime background by the curvature there \citep{wald1994quantum}. But one would also generally expect non-local corrections due to higher-order correlators in the quantized field, which may even dominate \citep{hu2008stochastic}. It is difficult to envision a general prescription for incorporating these corrections. Meanwhile, it is also unclear how one might proceed to associate (global) states in the QFT with local metric perturbations on some background --- as would presumably be sought on physical grounds for ``particle'' states modeled as particular spatiotemporally localized excitations about a Gaussian state.

The upshot of these difficulties is that it becomes natural to restrict attention to descriptions of very particular Gaussian states coupled to very specific spacetime backgrounds: backgrounds that are, up to a choice of cosmological constant in GR, appropriate global representations of vacuum in relativistic field theories (with, moreover, a trivial `pure gravity' sector) --- think: utterly empty spacetime, modulo cosmological constant \citep{selfreferenceEmptyspace}. These spacetimes are maximally symmetric, comprising a particularly well-behaved collection within the conformally flat members of the family of Einstein manifolds, or spacetimes whose Ricci curvature is given by a `cosmological constant' term (i.e. a term proportional to the spacetime metric).

Leaving the choice of cosmological constant unfixed in GR, one would like the quantized field prepared in such Gaussian states to moreover be invariant under the symmetries of the empty spacetime, as is deployed in the background. One thereby associates such Gaussian states that are, moreover, `maximally symmetric' as `zero-point' or (global) vacuum states of the quantized field. Hence, one concludes that the geometric structure of empty spacetime in the background is locally sourced by the \emph{zero-point energy} of that quantized field: the action of the `stress-energy tensor' operator on the quantized field under the prescription outlined above, \emph{specifically where the field is understood as prepared in its vacuum state}. Because of the symmetries involved, this is a quantity that must be constant on the spacetime: its role in semiclassically sourcing the spacetime geometry of the background therefore mimics the work done by a cosmological constant term in the classical theory GR. 

Focus now on the simplified case of a scalar QFT defined on a globally hyperbolic spacetime, and where suitable vacuum states are adiabatic or `Hadamard' --- roughly speaking, locally like the unique Poincar{\'e} invariant vacuum state in (special relativistic) QFT. This is a case of particular importance in the context of standard model cosmology. For instance, when the cosmological constant in GR is positive, empty spacetime sourced by the zero-point energy of the scalar field is uniquely best represented by de Sitter spacetime (cf. \citep{selfreferenceEmptyspace}, noting especially that elliptic de Sitter spacetime is not globally hyperbolic). In this case, there is a natural way of understanding that de Sitter spacetime background in terms of a Euclidean cosmos spatially expanding in cosmic time: i.e. uniformly dynamically stretching or scaling privileged families of Euclidean spatial coordinate frames, which are otherwise taken to be fixed along any one choice of moment within a `cosmic time' axis. 

Here is how this works \citep{schrodinger1957expanding}. Given an observer in the spacetime modeled as a maximal timelike (oriented) geodesic curve, the causal future (i.e. forward lightcone) of that observer may be foliated by a Euclidean spatial slice, which is thereby understood to exponentially expand in the observer's proper time. This construction leads to a `flat FLRW' frame on that observer's causal future, so that the restriction of de Sitter spacetime to any such region is often called the `flat FLRW patch' relative to them.\footnote{For more details on de Sitter spacetime and its many cousins, see \citep{Belotgoodtimesroll} and references therein --- especially \citep{schrodinger1957expanding} and \citep{calabi1962relativistic}. The `flat FLRW patch' is called the `cosmological patch' in \citep{Belotgoodtimesroll}, and the de Sitter metric within that patch can, in four spacetime dimensions, be written as the familiar flat FLRW metric $ds^2=-dt^2+e^{2t}(dx_1^2+dx_2^2+dx_3^2+dx_4^2)$. Although the flat FLRW patch is naturally understood in terms of the exponential expansion of space through a de Sitter observer's proper time, the intersection of the flat FLRW patch generated by that observer with the causal past of the same observer (i.e. the observer's backward lightcone) gives rise to the `static patch' relative to them. As the name suggests, within the static patch, the line element of the de Sitter metric can be written so that the spatial components are time-independent. This is clearly untrue with respect to the larger flat FLRW patch, and is as well untrue in de Sitter spacetime, considered globally.} Although neighborhoods of points within the flat FLRW patch relative to the observer retain the local symmetries of de Sitter spacetime, the flat FLRW frame there constrains the global symmetries relevant in the patch, recreating a symmetry group traditionally familiar in standard model cosmology. Namely: the patch may be interpreted by means of the flat FLRW frame in terms of a flat, infinite isotropic three dimensional space that eternally exponentially expands with the passage of a `cosmic time'. And importantly for below, with respect to the flat FLRW patch about the observer, space at any such moment in cosmic time behaves as a Cauchy surface. Meanwhile, the observer, on this interpretation, now belongs to an equivalence class of `stationary' (in other contexts, `comoving' with expansion) observers whose trajectories form a congruence within the patch. These are all the observers who bear witness to the exponential rate of expansion of a flat, infinite isotropic three dimensional space all around them. 

Nonetheless, note that with respect to the whole de Sitter spacetime, the initial observer is still distinguished as that particular member of this new family, about whom the FLRW patch is constructed. In what follows, I will refer to this initial observer as the \emph{focal observer}. Special care should be given to the fact that the choice of focal observer ``sticks around'' as a part of our description of the physics, so constructed. In a moment, this is what will allow me to define a canonical choice of shared (spatial) origin between `real space' and `momentum space' presentations of an EFT on the flat FLRW patch relative to the flat FLRW frame, so that a dynamics of rapid expansion of `real space' coordinates in cosmic time off of some initial moment are associated with a dynamical stretching of indexed fluctuations in `momentum space'. But it is important not to get ahead of myself! Nothing as of yet has brought me to the toolkit of EFTs.

Rather, so far I have considered a scalar QFT on de Sitter spacetime, whose zero-point energy one may now understand to source, as assessed from within the flat FLRW patch constructed about a focal observer, the exponential expansion of space in the background. And within that flat FLRW patch, for any moment in cosmic time, one may, centered in space about the focal observer, move freely between `real space' and `momentum space' presentations (with respect to the flat FLRW frame) of the QFT defined there. I turn now to the study of fluctuations therein.

Namely, in the `momentum space' presentation of the QFT, one may decompose the quantized scalar field in its vacuum state into two parts with the respect to the flat FLRW frame: a `homogeneous' component and perturbations thereof. The homogeneous component exactly satisfies the symmetries required of the spacetime background at each moment in cosmic time, and is therefore entirely well suited as an expression of the mean field in the QFT on that background, about which we might investigate the stability properties of the quantized field, with respect to the flat FLRW frame. That is, about the focal observer, given arbitrary perturbation modes introduced to linear order at a given moment in cosmic time all about them in space, one should like to model the growth of such modes in the vicinity of that moment. As a matter of physics practice, this is an invaluable procedure. Since the vacuum state is not an eigenstate of the field operator, there will be fluctuations about the mean field. In certain contexts determined by the relevant mean field physics, as in the present case (see \S2.3), those fluctuations may themselves be physically significant, giving rise to a stochastic ensemble whose back-reaction on the spacetime we should like to dynamically model. (And in such contexts within cosmology, where the cosmic time elapsed between two moments can be great, it is of even greater importance to track when and how long such effects of back-reaction might be safely ignored.)

Any such arbitrary perturbation modes defined along space about the focal observer at such an initial moment are indexed to frequency, as is standard in a momentum space presentation of a QFT. Less standard is the fact that those frequencies must themselves be defined in units that are matched to some characteristic spatial scale within the flat FLRW patch \emph{just at that moment in cosmic time}. The upshot is that any such perturbative modes indexed to their initial frequencies are dynamically decoupled from the transport of the characteristic scale through cosmic time that defines units in each moment. Hence, the modes dynamically stretch with the expanding background, so that those indexed to what are initially high frequencies `cross modes' to later resemble those that are then indexed to low frequencies. In this way fluctuations associated with perturbations in a scalar QFT relative to the flat FLRW frame act as `mode-crossing' degrees of freedom centered on the focal observer --- essentially like a system of decoupled harmonic oscillators belonging to that observer in that moment, which are produced in accord with their indexed frequency, and which thereby proceed to co-move with the exponential expansion of space all around. (Helpful diagrams may be found in \citep{brandenberger2013trans}.)

Finally, it is important to stress that this `dynamical stretching' behavior applies to vacuum fluctuations modeled perturbatively \emph{everywhere} in the flat FLRW patch. That is, one may associate with space at \emph{each} moment in cosmic time the production of such mode-crossing fluctuations that proceed to grow about the focal observer with the expansion of space, oscillating all the while --- at least in the vicinity of that moment. As I will return to in \S\ref{secEFTtoClassical}, these mode-crossing fluctuations --- when interpreted physically e.g. by means of an EFT toolkit (cf. \S\ref{secEFTcutoffs}) --- are what ultimately render inflationary cosmology a ``causal mechanism for generating'' (e.g. \citep[p. 177]{brandenberger2000inflationary}) the primordial perturbations that are ultimately responsible for seeding the classical evolution of large-scale structure observed today. 

\subsection{EFT cutoffs}\label{secEFTcutoffs}

To stress a virtue of the presentation so far: note that, to this point, I have not --- nor have I had any need to --- make use of an EFT toolkit. In this respect, the dynamical stretching in cosmic time of vacuum fluctuations in a perturbative, frame-relative treatment of a scalar QFT on a flat FLRW patch relative to a focal observer in the nicely behaved de Sitter spacetime \emph{is a feature of semiclassical gravity simpliciter}, as would be relevant to a certain modeling context. That is to say, it follows (merely) as a consequence of our means of amalgamating, by means of QFTCS, GR and QFT descriptions of certain classical gravitationally coupled material quantum systems prepared in especially well behaved states. Meanwhile, it just so happens that those systems could be relevant in a fundamental physical modeling approach to large-scale cosmology.

\emph{On the other hand, the physical interpretation of these vacuum fluctuations in such a perturbative treatment has not yet been provided.} And so, there is little reason yet to care that such fluctuations would grow with expansion. In this respect, I have not yet rendered out of the process of amalgamation of current physics just summarized any approximate description that is anticipated of the low-energy physical consequences of the future UV theory of quantum gravity applied to such a system, in light of the perturbative mode-crossing behavior just documented. To accomplish such a further aim, it is customary to turn to an EFT toolkit, which is ordinarily used in interpreting QFT within the context of standard model particle physics. 

As a brief aside, but one which merits more than a footnote: I take this theoretical commentary about when exactly an EFT toolkit finally enters to do interpretive work as clarifying --- in the form of providing explicit grounds for --- the view expressed by \citet{brandenberger2021limitations} that trans-Planckian censorship presents, specifically, a problem in our applying EFT thinking to cosmology. The flip-side of this is that, e.g., inflationary dynamics may indeed turn out to be an effect of UV quantum gravity under some notion of low-energy description other than that provided by an EFT toolkit.\footnote{How this would go is, unfortunately, not immediately clear. In particular: this other notion of low-energy description would, presumably, also be tasked with explaining why the EFT toolkit is, in nearly every (other) context, sufficient for what is demanded of it. One suggestion, motivated by an objection to trans-Planckian censorship developed in \citep{dvali2020inflation}, is to appeal to a more rigorous notion of scale decoupling in fundamental physics, which might recover the EFT toolkit in the latter's more ordinary applications. One notable challenge, in such a case, would be locating independent support for that particular rigorous notion in the context of UV quantum gravity beyond field theory.} This, then, circles back to the point I make in \citep{selfreferenceTransPlanckianPhilosophy} that an embrace of trans-Planckian censorship \emph{together with a sustained commitment to inflationary dynamics in early universe cosmology} may render early universe inflationary structure formation an instance of UV quantum gravity phenomenology. Namely, in contrast with traditional thinking about inflation, inflationary dynamics \emph{with trans-Planckian censorship precluding an inflationary EFT} presents as an empirical strain of research, which is non-trivial to recover in the course of developing a future empirically adequate UV quantum gravity theory.

Not to get waylaid by the philosophical foundations of EFTs, it suffices here to rehearse the standard line. In the context of (special relativistic) QFT, treating a given constructed theory as an EFT amounts to specifying an energy scale cutoff, only below which is the theory taken to be accurate of a given physical system, precisely as it is written. And although great care must be given to how one proceeds without breaking global Poincar{\'e} invariance in the theory (that is, below the cutoff), it is nonetheless standard to treat the units about which the cutoff is defined in momentum space in terms that correspond to coordinates declared along some specified Cauchy surface embedded in the Minkowski spacetime background (usually given by spatial projections in an ordinary inertial coordinate frame on that background).

This EFT toolkit is helpful for insulating the descriptive accuracy of our current standard model particle physics within accessible empirical regimes from radical theory change, in the pursuit of a new means to model the same physical systems in more exotic regimes beyond our current empirical access --- the paradigmatic case being a regime in which there is sufficient energy to produce some never-yet-observed, very heavy massive particle. In the context of gravity, various heuristic arguments have been put forth to suggest that we ought to identify the Planck scale as one obvious choice of where to place an upper limit for \emph{any such UV cutoff}: the idea being that, optimistically, QFT constructions of some or other sort will remain descriptive of gravitationally coupled material quantum systems, exactly as written, up until we might probe near the physical `Planck regime' \citep{hossenfelder2013minimal}.

Apropos the discussion in the previous subsection (see also \citep{koberinskiSmeenkRethinkingCCP}), it must be that the units against which a UV cutoff is defined, in the case of QFTCS, are specified relative to some salient extra structure on the spacetime background. In our particular case of interest, one proceeds by defining those units with respect to space about the focal observer, as defined in terms of an arbitrary choice of moment in cosmic time. So, the Planck scale, delimiting the physical Planck regime about the focal observer in spacetime --- as relevant to treating the QFT on that spacetime as an EFT there --- is moreover defined as relative to a choice of moment in cosmic time. 

Note that, proceeding in this way, one \emph{finally utterly discards} the global hyperboloid geometry of de Sitter spacetime, in favor of treating as physical the global structure given just by the flat FLRW patch, explicitly foliated by a Euclidean Cauchy surface that expands in cosmic time --- i.e. rendering as physical content the flat FLRW frame on the patch. In other words, the move to treating the QFTCS construction as an EFT, in virtue of the physicality of the UV cutoff therein along a Euclidean choice of Cauchy surface, would seem to assign genuine physical content to what was, thus far, only a formal decision to focus on the causal future of the focal observer in some larger spacetime, with the help of some auxiliary formal structure therein. By contrast, now with respect to any observer within the patch, one may distinguish the physical `Planck regime' about them at any moment in cosmic time as a sufficiently small neighborhood centered on them in the relevant Euclidean Cauchy hypersurface. Within this regime, for each such observer, it is entirely inappropriate to take descriptions provided by the QFTCS construction to be physically apt. And in the case of the family of stationary observers who co-move with the focal observer and bear witness to exponentially expanding space, this Planck regime is, in any such moment in cosmic time, given as a solid metric sphere --- the surface of which describes the exit from the Planck regime, or a surface at which physical descriptions provided by the QFTCS construction are abruptly taken to become apt. 

As I discuss in \citep{selfreferenceTransPlanckianPhilosophy} in connection with an approach to the trans-Planckian problem in cosmology developed by \citet{brandenberger2013trans}, this surface in space gives rise to a timelike hypersurface in the background spacetime, which one may understand to radiate trans-Planckian signals into the spacetime relative to the observer, spoiling the sense in which space about that observer in any given moment is, in fact, Cauchy. But these signals do not themselves come from the EFT, whose UV cutoff precludes our physically interpreting vacuum fluctuations in the quantized field that would radiate across the exit from the Planck regime. Meanwhile, vacuum fluctuations in the quantized field below that cutoff may be associated with dynamically stretching, independent oscillatory modes produced along space about the focal observer over scales that do not resolve the exit from the Planck regime as a surface distinct from the ideal observer at the Planck regime's center.

That is the physics to do with the UV cutoff in an EFT treatment of the quantized field in its vacuum state on the exponentially expanding background. But there is, arguably, a second cutoff to countenance: a low-energy, `infrared' (IR) cutoff at the other end of the energy spectrum. Discussing this second cutoff will close out the subsection. 

Except perhaps in the context of cosmology, the IR cutoff is often forgotten with little consequence --- the quantized field relevant for some given non-cosmological application is typically understood as `placed in a finite (perhaps unfathomably large) box'. Nonetheless, it is a well-understood feature of an EFT toolkit, applied to quantized fields as a means of physically interpreting fluctuations indexed by frequency in a perturbative momentum space presentation of the QFT. Plausibly, the cutoff distinguishes a frequency scale beyond which some feature of the underlying physics that is otherwise aptly described in terms of a quantized field in fact undermines the use of a linear perturbation theory there. 

In the context of standard model cosmology, this perspective leads one to identify the Hubble radius as one natural choice of an (inverse) frequency scale at which to place the cutoff. But this is just one choice of where to place the cutoff --- one which is likely made less compelling by the fact that the Hubble parameter used to fix that upper bound is itself not constant in cosmic time, suggesting a sort of `decoupling' of the Hubble constant from any particular underlying fundamental physics. Other choices of IR cutoff have also been explored in the context of quantum gravity research, especially in the context of holographic dark energy proposals (see, e.g., \citep{cohen1999effective} and \citep{li2004model}). There are subtle arguments behind these various proposals that have to do with `UV/IR' mixing. Still, for the present purposes of clarifying the TCC, it will suffice to stick to the former proposal: an IR cutoff opposite the UV Planck scale that is drawn on the order of the (inverse) Hubble radius. There is some extrinsic motivation to support this decision at least in the context of inflationary applications, as I will now discuss.

\subsection{An `EFT-to-classical' transition}\label{secEFTtoClassical}

What happens when fluctuations are `trans-Hubble', crossing from physically permissible modes to modes beyond the specific IR cutoff just put forth? As noted originally by \citet{polarski1996semiclassicality}, given a mandate that the application of the linear perturbation theory ought to be approximately metastable \emph{far off of any suitably chosen Cauchy surface}, it would seem to be the case that trans-Hubble modes \emph{effectively classicalize}: freezing out as classical perturbations that, on average, obey the symmetries of the background spacetime --- essentially as overdamped oscillators. Disucssing this effective form of classicalization will be the final installment in the contemporary anatomy of the TCC.

Here is how it goes. Stability arguments are a mainstay of modern physics \citep{fletcher2020principle}, and approximate metastability is often treated as a weakening of the goal of self-consistency in physical description. In this particular case, recall that QFTCS is intended, in the first place, to describe the behaviors of quantized fields on arbitrary spacetime backgrounds. This is a fully four-dimensional, `timeless' project. But in the previous subsection, I associated with the quantized field a persistent, constant dynamical production of physically significant fluctuations across space, as expands about the focal observer in their proper time. Conceivably, such processes occurring at early stages might interfere with the approximate descriptive relevance of those same processes associated with later stages. Understandably, one would like to preclude this; the mandate flagged in the previous paragraph stipulates that this be so. 

More precisely, what the mandate justifies is that, across space at each moment in cosmic time within the FLRW patch, one may decompose each physical fluctuation mode therein, restricted to that surface, into two parts: a component that contributes to a classical ensemble satisfying the symmetries of the background spacetime at super-Hubble length scales and an `excess' component that encodes small-scale, symmetry-breaking behavior. What motivates this particular mode-by-mode decomposition, given the mandate? Cosmic time-translations are symmetries of the particular spacetime background, and so the zero-point state of the field defined there should likewise be invariant. Having split off the homogeneous component of that state --- itself invariant under isometries of the background --- one ought as well to regard the perturbative component as invariant, at least at some level of approximation or course-graining over which we are interested in the metastability of the system. 

The Hubble radius places a rough upper bound on the causal connectibility of local, microphysical processes idealized as spacetime events co-occurant along space in a given moment, treated as a coarse-grained, classical description of physics. It is then a feature of the classical theory of general relativity, treated as a local gravitational theory that is descriptive of low energy perturbations against an exact solution to the theory that serves as background, that we may \emph{ignore} super-Hubble perturbations on such a background, in doing so. Or rather: sub- (inverse) Hubble frequency contributions may be absorbed into the description of that background, against which super- (inverse) Hubble frequency gravitational physics might be perturbatively assessed. Meanwhile, the excess components within the super- (inverse) Hubble fluctuations oscillate and rapidly dampen. So, only the sub- (inverse) Hubble quasi-symmetric modes that comprise the above ensemble remain as squeezed contributions to the background. The upshot is that we may treat the `small-scale' excess components as, mode by mode, the environment to which the metastable, sub- (inverse) Hubble components are coupled. Tracing over that environment, one may expect, comoving with expansion, rapid decoherence in the sub- (inverse) Hubble components that each satisfy, on average, the symmetries of the spacetime background. 

This process is typically understood as an instance of `decoherence without decoherence' (e.g. \citep{polarski1996semiclassicality}), following Wheeler's perspective on environment-induced decoherence in open quantum systems. The key point here is just that, although decoherence is not strictly achieved (there is no mechanism that would warrant severing off the small-scale `excess' degrees of freedom as environment), something else like it is --- namely, the effective classicalization of \emph{just the large-scale degrees of freedom we care about in cosmology} (for some discussion of this focus in cosmology, see \citep[\S1]{vidotto2017relational}). Expectation values of the physically relevant quantities in the large-scale application of this vacuum EFT --- namely, the amplitudes of the metastable, sub- (inverse) Hubble components of the fluctuation modes --- can be calculated as one would classical stochastic observables. Amongst such classically-behaving field degrees of freedom, those that ever dynamically fall back, now as classical degrees of freedom, to sub-Hubble length scales can ultimately seed `large-scale structure' relevant at later stages --- classical linear perturbations defined on the expanding background.

As a final remark, note that despite the intonation `large-scale structure', nothing in the above physics depends sensitively on our current thinking about our cosmos. This `EFT-to-classical' story is a feature of such QFTCS constructions on rapidly expanding backgrounds whose sub-UV cutoff fluctuation modes in a suitable perturbation theory are physically interpreted by means of an EFT toolkit. In this respect, an inflationary hypothesis in the context of the early universe to explain downstream classical structure formation is merited by just appeals to our current physics: a quantized inflaton field, treated like any other such field in contemporary particle physics, would yield dynamics in the early universe that are empirically relevant, given background commitments in large-scale cosmological modeling. 

On the other hand, that such a proposal should yield dynamics in the early universe that are empirically relevant, given those background commitments, requires suitable initial conditions in our current thinking about our cosmos. Namely, one supposes at the onset of inflation that there are no populated sub- (inverse) Hubble modes, so that everything that classicalizes through inflation comes, initially, from within it, i.e. as vacuum fluctuations. This assumption is easily won in the case of setting the IR cutoff at the (inverse) Hubble scale. But other choices of IR cutoff, e.g. motivated by speculations about UV/IR mixing in quantum gravity research, may apply legitimate pressure on the assumption. 

\paragraph{}
So concludes a contemporary anatomy of the conjecture.

\section{Why censorship?}

Apropos the anatomy just developed, in certain background expanding spacetimes, i.e. many of those familiar in inflationary cosmology --- as well as in future-asymptotically de Sitter scenarios (like our own standard model) --- physically salient quantum fluctuations can dynamically stretch to IR, sub- (inverse) Hubble scales. These stretched modes thereby classicalize, ultimately becoming relevant to the classical physics of large-scale structure formation and growth in our cosmos. In such cases, if the background expansion of space lasts sufficiently long, the empirical record of large-scale structure at sufficiently late stages would carry trans-Hubble imprints of specifically trans-Planckian modes:  arbitrarily high-energy physics, which is associated with the physical Planck regime of the quantized field along some early Euclidean Cauchy surface, say at the onset of inflation. 

But, of course, EFTs are theories that are not to be trusted about physics above their high-energy cutoffs, i.e. as defined along such Cauchy surfaces. How, then, are we to make sense, in our cosmos, of such empirical imprints of artifacts in the theory along the earlier Cauchy surface, which are themselves not intended to be interpreted physically? The TCC dodges the question: there are, by virtue of the future quantum gravity theory apt for the Planck regime about our focal observer, no such imprints to be noted. Rapid expansion \emph{simply cannot last long enough} to dynamically stretch trans-Planckian modes at the onset of inflation into trans-Hubble ones, as would then be suited for classicalization and observation. 

As just stated, it is perhaps easiest to understand the TCC as a methodologically conservative patch on a particular problem. Namely, per the TCC, we can disregard cosmological scenarios in which the relevant spacetime background would lead to such troublesome empirical imprints of trans-Planckian physics. But methodological conservatism is a thesis about holding onto our current scientifically informed beliefs for as long as is viable, absent new reason to discard them \citep{sklar1975methodological}. And except for our belief in inflationary dynamics within the early universe, we would seem to have no current need to patch any such problem. In this sense, the TCC would seem to be a methodologically conservative patch on a particular \emph{cosmological} problem --- and specifically \emph{within} inflationary cosmology. Yet it is embraced at a rather severe cost of speculating entirely new physics relevant in the ongoing development of a theory of UV quantum gravity. 

To what extent is the cost imposed too great --- or, in other words, to what extent is the speculation sufficiently radical as to undermine the spirit of methodological conservatism embodied in its initial proposal? The spectacular advertisements rehearsed in the Introduction suggest that there is basically no cost: that it is hardly radical to embrace such a speculation. In other words, these advertisements of the TCC concern our present thinking about what will \emph{inevitably come} of a future theory of UV quantum gravity. Beneath the promotional level, they are claims made today about how we reckon UV quantum gravity \emph{will work}. In light of the anatomy just provided, I am now in a position to clarify what lies at the heart of such claims.

Namely, the new physics asserted in the TCC is that UV quantum gravity pairs together the IR cutoff in effective field theoretic descriptions of quantum gravitationally coupled material quantum systems with the scale implicated in effective classicalization of quantum fluctuations --- specifically, as the latter arises in semiclassical modeling relevant within large-scale cosmology. And this strikes me as somewhat mysterious: in order to meaningfully wield trans-Planckian censorship in early universe inflationary cosmology, there must be some fated harmony between fundamental quantum physics and the large-scale semiclassical cosmos (that is, the cosmos, defined effectively with respect to all small-scale degrees of freedom in our universe serving as environment). Supposing that the IR cutoff is much larger than the Hubble radius at the onset of inflation, one might then regard the disavowal of any initially excited sub- (inverse) Hubble modes as an alarming case of fine-tuning. Yet, on the other hand, supposing that the IR cutoff is on the same order as proposed, one is reminded in trans-Planckian censorship eerily of Eddington's fated attempts to tie the value of the cosmological constant and the length scale of the entire closed classical Einstein universe at the start of expansion to the mass-to-charge ratio of the electron in high-energy quantum physics \citep{kilmister1994eddington}. This does not amount to an objection to conjecturing trans-Planckian censorship, but it does seem as if it might give one pause.

\section{Conclusion}

As I have already argued in \citep{selfreferenceTransPlanckianPhilosophy}, the TCC stipulates that unknown Planck-scale fundamental physics (so, UV quantum gravity phenomenology) must be physically screened off from our cosmological descriptions of structure formation and growth.  This is  particularly  pressing  in  the  context  of  inflationary  cosmology  in  the  early universe, but as mentioned in the Introduction, it has also been used in the context of the cosmological constant in late-stage cosmology beyond the standard model. As I elaborate in that article, this state of affairs might reasonably be seen as exciting, influencing its general uptake as a topic of study within our thinking about semiclassical gravity.

Still, one might worry that the work done by trans-Planckian censorship in cosmology is insufficient to assuage concerns that such a conjecture regarding fundamental physics  has ultimately come out of thin air. In this case, the spectacular advertisements for the TCC rehearsed in the Introduction are likely to be encouraging: trans-Planckian censorship may almost be considered inevitable as a feature of the future UV quantum gravity theory, in which case the problem it would patch we may look forward to being patched `for free' by any such adequate theoretical achievements. 

In this article, I attempted to provide a presentation of trans-Planckian censorship exclusively in terms of our current, well-trodden physics of quantized fields, spacetime, and (classical) gravity in order to better assess the argumentative merits that ultimately underpin those advertisements --- to judge how costly is the TCC as a speculation specifically concerning low-energy cosmological applications of a future UV quantum gravity theory. On this front, it is hopefully clear, in the first place, that the conjecture has very little to do with early universe cosmology, \emph{except were we to there hold onto inflationary dynamics as the causal mechanism for the downstream formation of cosmic structure}. Namely, we can understand the consequences of the TCC entirely in terms of our well-trodden physics, so that --- given the particular conjectured further claim about UV/IR mixing in UV \emph{quantum} gravity --- we may ask whether certain applications of our current fundamental physics require a physical interpretation other than that which makes use of an EFT toolkit. That is to say, in early universe cosmology, it is not the case that we may reason from a conjectured trans-Planckian censorship to a conclusion that the application does not occur \emph{which would otherwise be made complicated by censorship}. This is contrary to what is suggested in \citep{bedroya2020trans}: that in virtue of the TCC applying pressure on inflationary cosmology, we might turn to consider alternatives instead. (Alternatives may be sought on independent merits, whose embrace would render trans-Planckian censorship impotent in early universe cosmology. But this is another matter --- perhaps relevant in \emph{confirming} candidate quantum gravity theories, where they happen to give rise to such explicit alternatives.)

Meanwhile, we may also understand there to be substantive new fundamental physics contained in any affirmation of the TCC. Consistency of our as-yet unknown UV theory of quantum gravity ensures that physical applications of EFTs involving an expanding spacetime background come equipped with bounds, owing to whatever in that future theory grounds the UV and IR scales that show up in the formalism suited for providing such effective descriptions. So we may be wary of the employ of trans-Planckian censorship in proposals like that due to \citet{ijjas2019new}, which otherwise brand themselves as proposals that successfully insulate descriptions of large-scale cosmology from any unknown UV quantum gravitational physics. In fact, one might argue that the standard model in late-stage cosmology is \emph{already} successfully insulated from any unknown UV quantum gravitational physics. Meanwhile, any leap from consideration of that standard model to a cyclic model beyond it would seem to require invoking a principle like trans-Planckian censorship (cf. footnote \ref{fnCyclic} above) --- spoiling the claim of insulation. Hence, as in Eddington's program in the early history of relativistic cosmology, one sees here a new effort to render the large-scale universe in its entirety as a high-energy quantum phenomenon: that is, as a readily (i.e. classically) observable empirical consequence of certain physical facts concerning UV quantum gravity. 

Finally, on this last point, I will reiterate --- now in a different light --- concerns raised above about the harmony between the IR scale relevant in the fundamental physics posited about the future UV quantum gravity theory and the scale relevant to effective classicalization in cosmology. Namely, it seems unlikely, in the first place, that this harmony would be secured merely by analogy to renormalization in the case of quantum gravity beyond field theory, even if that argument is --- as the reasoning in \citep{bedroya2020sitter} suggests --- sufficient to establish that there is \emph{some such IR cutoff} to countenance. On the other hand, the comparison to cosmic censorship may be particularly apt in regards to this concern: so too in the case of cosmic censorship, there is some interplay posited between quantum processes `near' a timelike singularity and the classical physics that emerges as relevant sufficiently far away in spacetime, i.e. beyond some horizon. But cosmic censorship is many things at once \citep{earman1992cosmic}, and taken conservatively it describes a decades-old program of research in general relativity that has very rarely been connected explicitly to the vanguard of quantum gravity research. More work on this latter comparison is thus, I think, merited for the sake of evaluating the prospects of trans-Planckian censorship as an avenue of research that concerns our understanding of the conceptual structure of the problem of quantum gravity, especially in relation to the swampland.

\bibliographystyle{abbrvnat}
\bibliography{transplanckiancensorship}

\end{document}